\documentstyle[11pt,Moriond,epsfig]{article}

\def\Journal#1#2#3#4{{#1} {\bf #2}, #3 (#4)}

\def\NIM{\em Nucl. Instrum. Methods}

\def\NP{{\em Nucl. Phys.}}

\def\be{\begin{equation}}
\def\ee{\end{equation}}
\def\bea{\begin{eqnarray}}
\def\eea{\end{eqnarray}}

\begin{document}
\vspace*{4cm}
\title{RECENT RESULTS ON STRANGENESS PRODUCTION FROM CERN EXPERIMENT NA49}

\author{D. VARGA \emph{for the NA49 Collaboration}}

\address{Roland E\"otv\"os University, Department of Atomic Physics, \\
	1 P\'azm\'any P\'eter S\'et\'any, H-1117 Budapest, Hungary}

\maketitle\abstracts{New results from NA49 on the production of
	strangeness in elementary and nuclear reactions are
	presented. New measurements of charged kaon and pion
	production have been obtained from 40 AGeV Pb+Pb
	collisions. The evolution of strange meson yields in
	nucleus-nucleus interactions is studied as a function of
	collision energy and system size. Cascade baryon yields are
	presented for a wide range of hadronic interactions including
	first measurements in p+p and centrality controlled p+Pb
	collisions.}

\section{Introduction}

%\begin{figure}[!ht]
%\psfig{figure=experiment.eps,height=2.5in}
%\caption{Experiment layout \label{fig:layout}}
%\end{figure}

	NA49 is a large acceptance hadron spectrometer, which is
	designed to explore a wide range of hadronic interactions at
	different beam energies. The tracking system based on large
	TPC's allows a systematic study from elementary collisions
	such as p+p and $\pi$+p through the more complex p+A
	interactions to A+A interactions. A dedicated counter was
	designed to control the centrality of p+A collisions, whereas
	the system size dependence of A+A collisions can be studied by
	measuring C+C and Si+Si collisions from fragmentation
	beam. The experimental setup is described in detail in
	\cite{nimna49}.

\section{Strange Meson Yields in Nucleus-Nucleus Collisions}

	The main motivation to study A+A interactions is the search
	for the deconfined state. A signature of the onset of
	deconfinement would be the increase of the produced
	strangeness, measured by the ratio of kaons to pions.

%\subsection{New Results in Pb+Pb Interactions at 40 AGeV beam}

%	New results, kaons and pions, independent methods

%\begin{figure}[!ht]
%\psfig{figure=k+.ps,height=2.5in}
%\psfig{figure=k-.ps,height=2.5in}
%\caption{Kaon and pion spectra at 40 GeV \label{fig:result40}}
%\end{figure}

%	Results of 4pi yields for K and pi

\subsection{Beam Energy Dependence}

	New results have been obtained in central Pb+Pb collisions at
	40 AGeV beam energy for kaon and pion multiplicities
	\cite{qmblume}. Two independent analysis methods based on time
	of flight and specific energy loss ($dE/dx$) measurement are
	found to be consistent. The total yield per event measured in
	the 7\% most central collisions is $56.3 \pm 3$ for positive
	and $17.8 \pm 0.9$ for negative kaons.

\begin{figure}[!ht]
\begin{center}
\psfig{figure=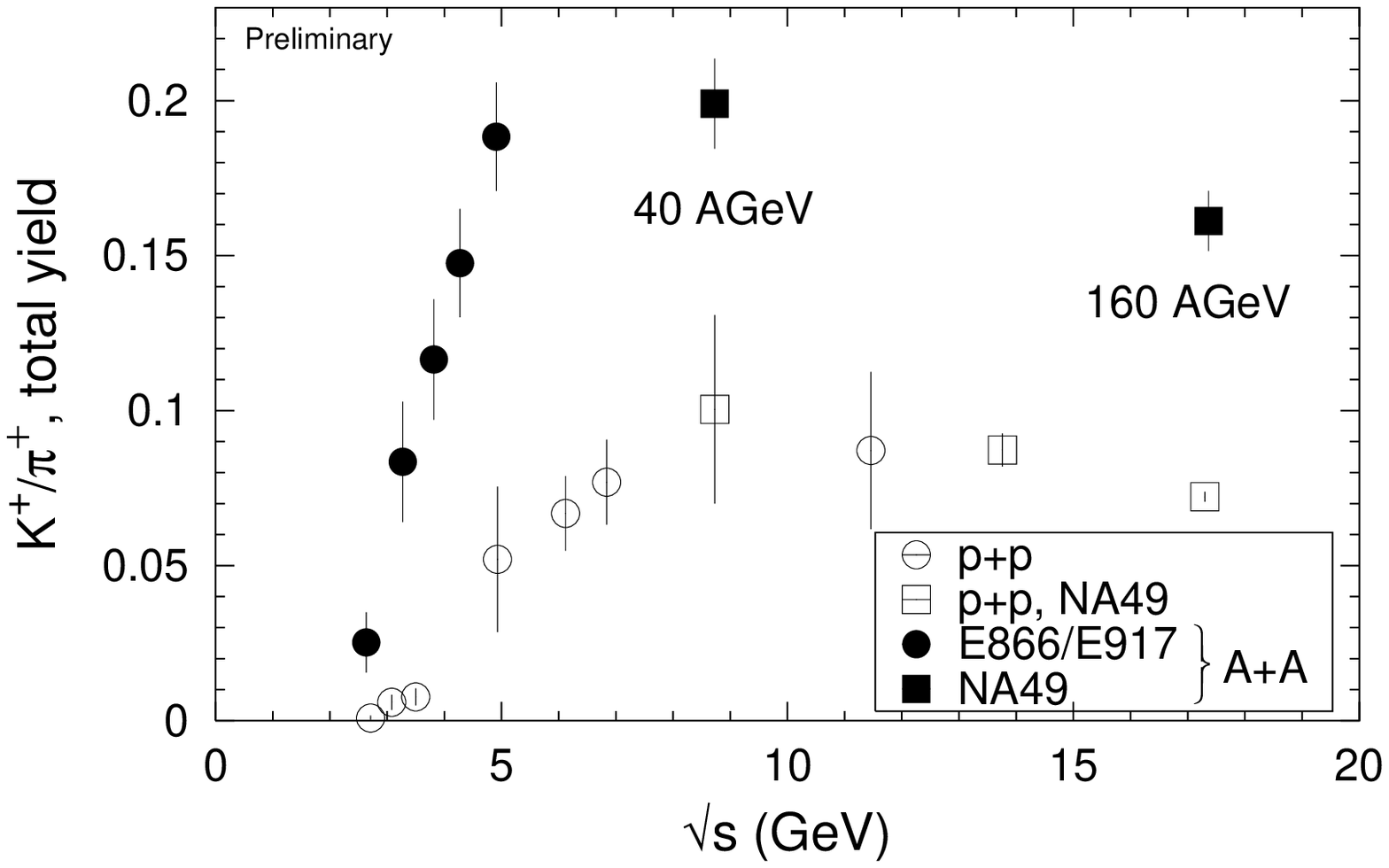,width=2.6in}
\psfig{figure=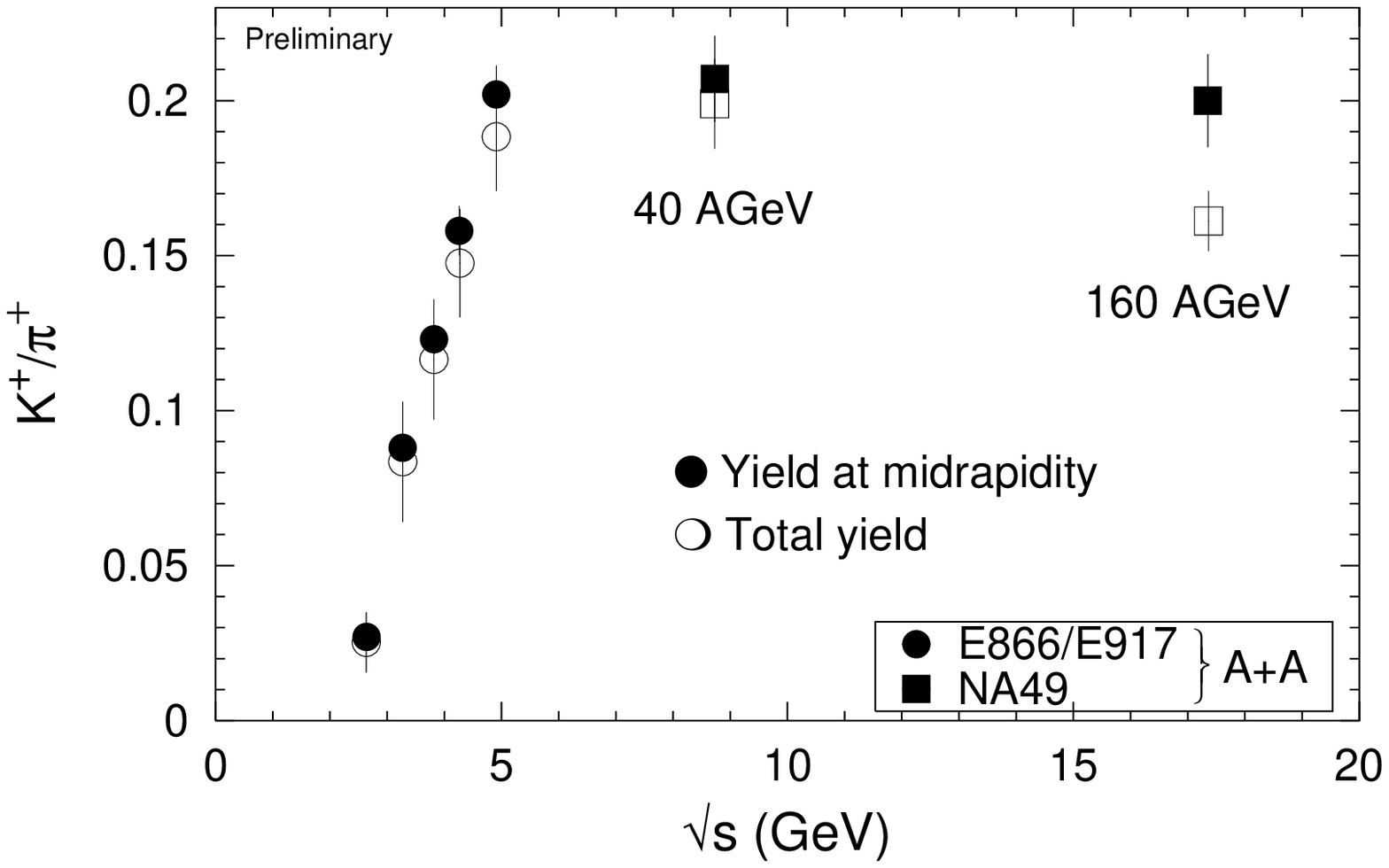,width=2.6in}
\end{center}
\caption{$K^+/\pi^+$ as a function of collision energy. ({\em Left:}) The ratio of
total yields in A+A and p+p collisions. ({\em Right:}) Comparison of
the ratio of total yields and midrapidity yields
\label{fig:kperpi}}
\end{figure}

	The energy dependence of the $K^+/\pi^+$ ratio is shown in
	Figure \ref{fig:kperpi}. The ratio of the total yields
	increases at AGS energies \cite{ags1}, %\cite{ags2}
	shows a maximum around the 40 AGeV point and decreases slightly
	towards the top SPS energy. For comparison, the $K^+/\pi^+$
	ratio in p+p collisions is also shown. The nonmonotonic
	behaviour is less pronounced for the yields measured at
	midrapidity. The analysis of last year's 80 AGeV beam and next
	year's 20 and 30 AGeV datataking will clarify the behaviour.

\subsection{System Size Dependence}

\begin{figure}[!ht]
\begin{center}
\psfig{figure=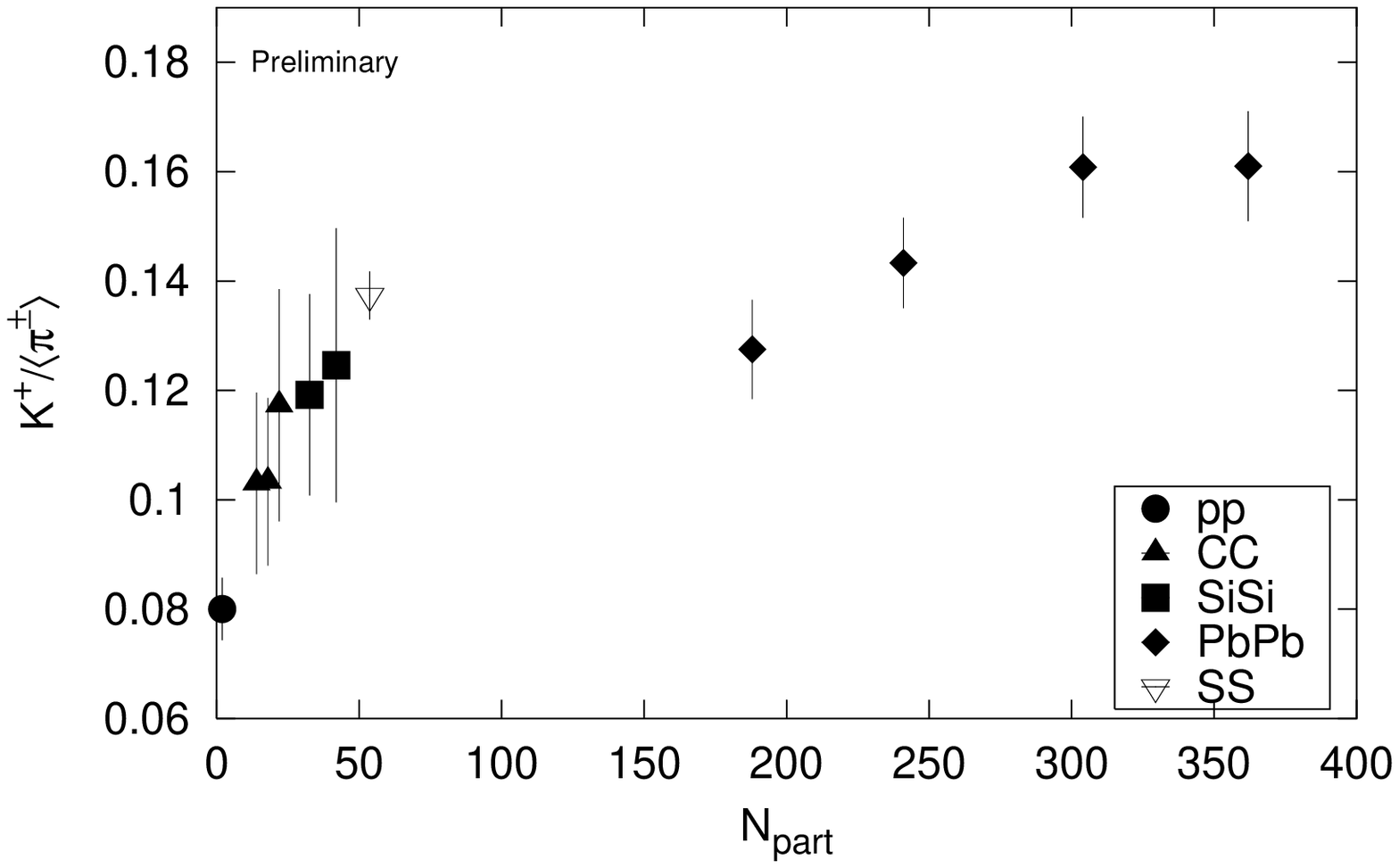,width=2.6in}
\psfig{figure=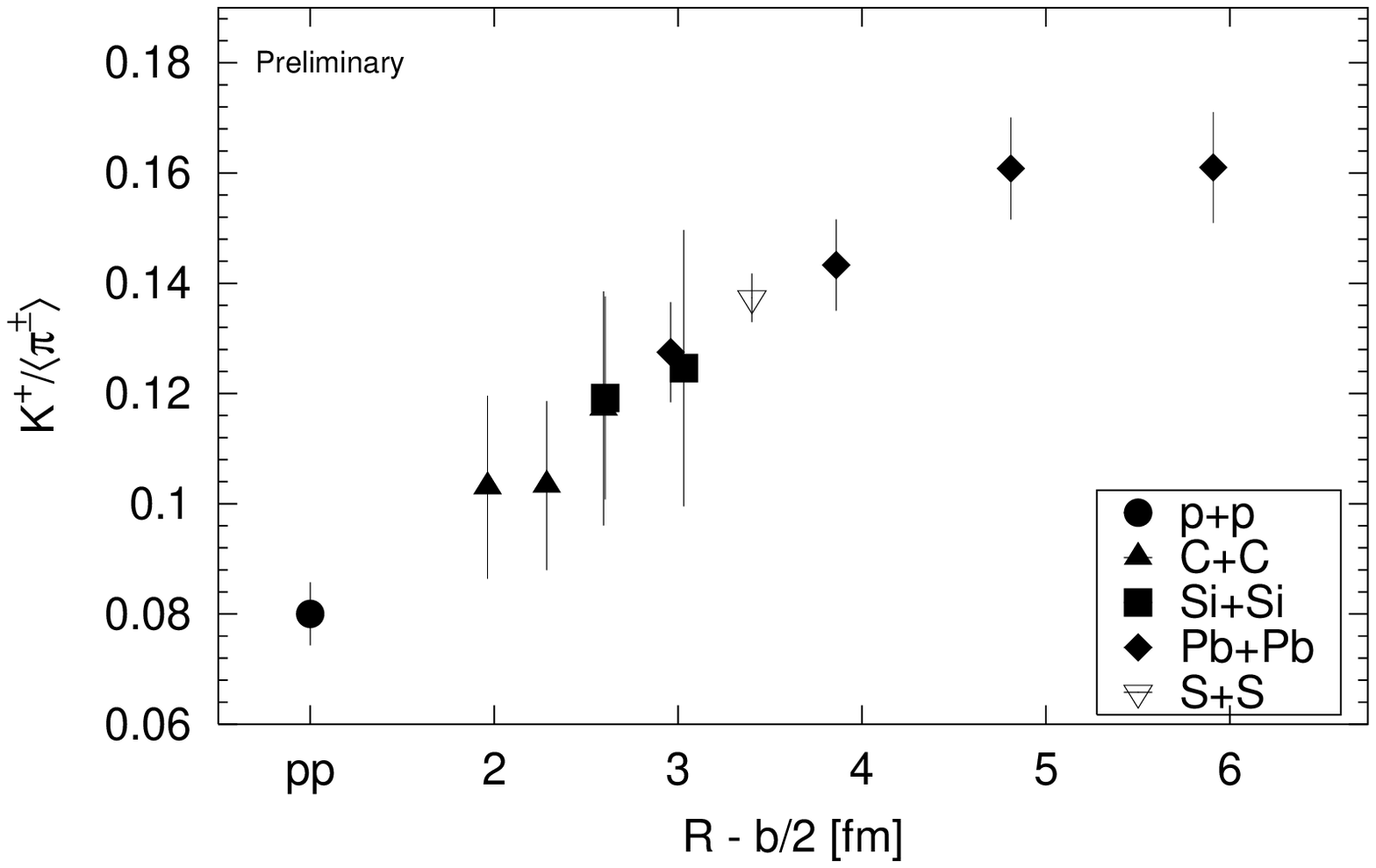,width=2.6in}
\caption{$K^+/\pi^+$ ratio at different system sizes, versus
$N_{part}$ ({\em left}) and versus $R-b/2$ ({\em right}) \label{fig:thick}}
\end{center}
\end{figure}

	In the search for the deconfined state, the other relevant
	initial parameter besides the beam energy is the size of the
	colliding nuclei. To study this dependence, C+C and Si+Si
	collisions were measured: Figure
	\ref{fig:thick} shows the $K^+/\pi^+$ ratio compared to Pb+Pb
	and S+S measurements.

	The results clearly show that the number of participants
	($N_{part}$) does not allow a common scaling for different
	system sizes and centralities. Instead of $N_{part}$ another
	scaling variable can be invented \cite{ismdsikler}: plotting
	against $R-b/2$, where $R$ is the radius of the nuclei and $b$
	is the impact parameter, the points fall on a uniform
	curve. This scaling variable corresponds to the width of the
	interaction zone. The analysis of the full available
	statistics will decrease the errorbars.

\section{Cascade Baryon Production}

	The cascade baryons ($\Xi^-$ and $\bar \Xi^+$) carry double
	strangeness, so they are more sensitive to strangeness
	production. The NA49 tracking system allows the measurement of
	these weakly decaying particles via their decay products
	\cite{barnby} \cite{barton}. The acceptance extends over a
	wide rapidity range ($\pm 1$ unit around midrapidity) and full
	transverse momentum range, from $p_T=0$ up to 2 GeV. The
	centrality of p+A collisions is determined with a dedicated
	counter which detects the ``grey protons'', the recoil protons
	from the target. From that, using the VENUS model, the number
	of collisions ($\nu$) can be determined. The centrality
	counter is also used for online triggering to enrich the more
	central sample.

\subsection{Cascade Baryon Yields in Proton-Proton and Proton-Lead Collisions}

\begin{figure}[!ht]
\psfig{figure=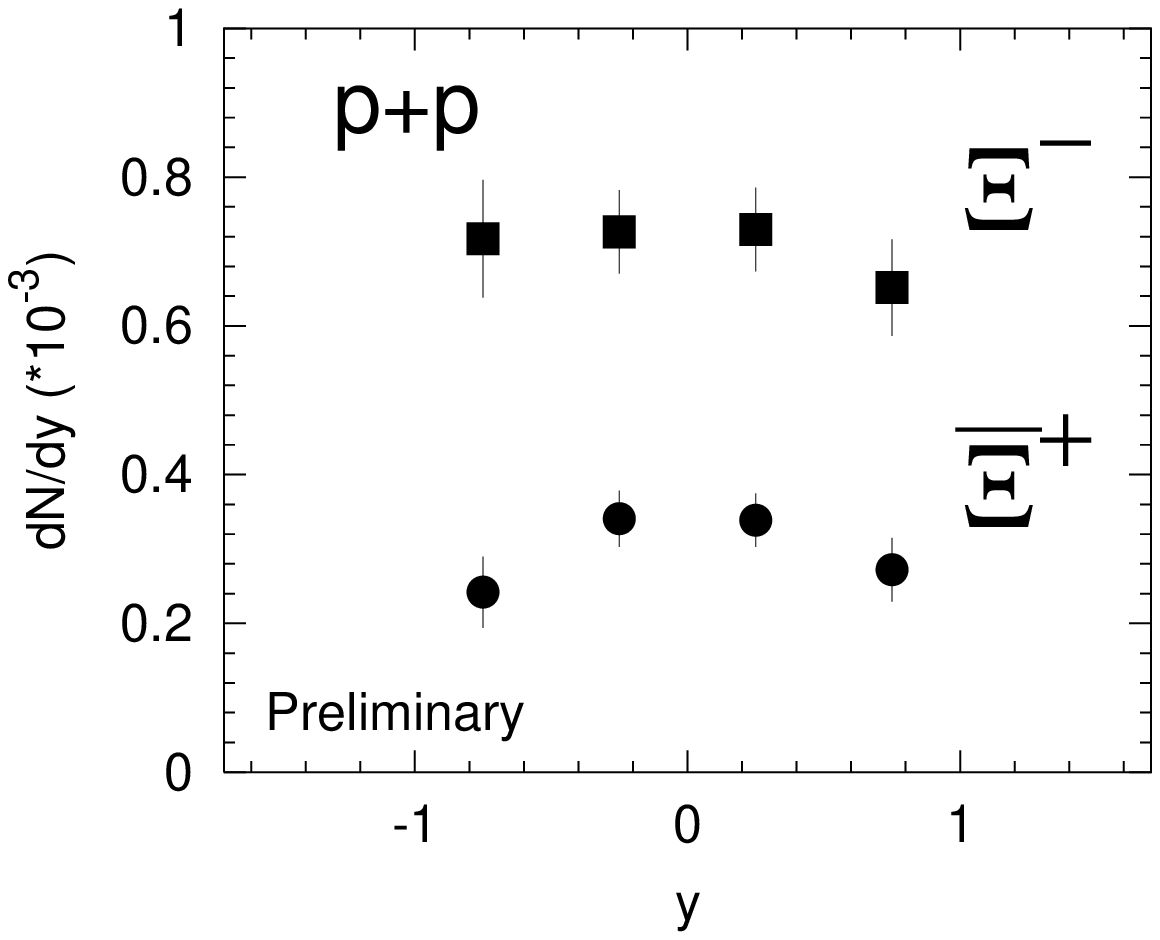,width=2.0in}
\psfig{figure=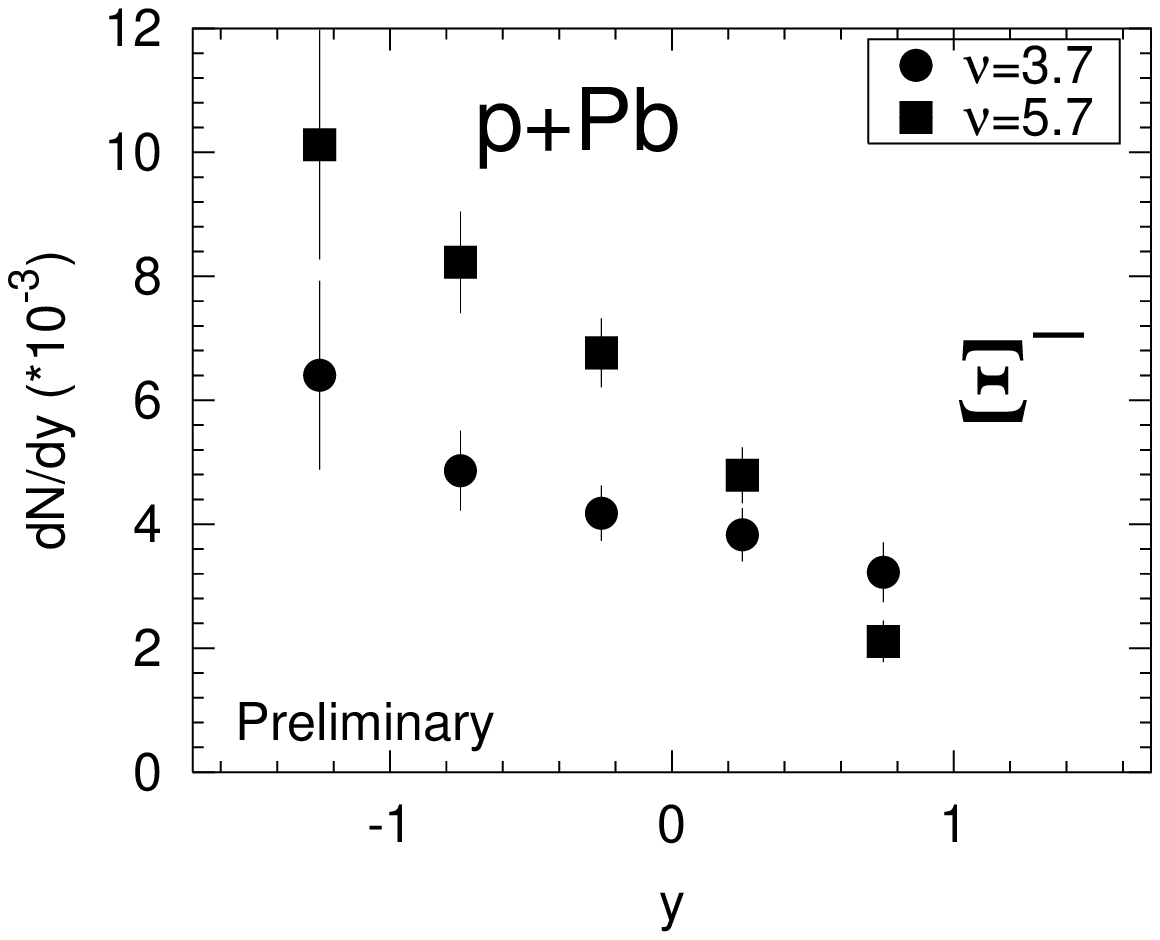,width=2.0in}
\psfig{figure=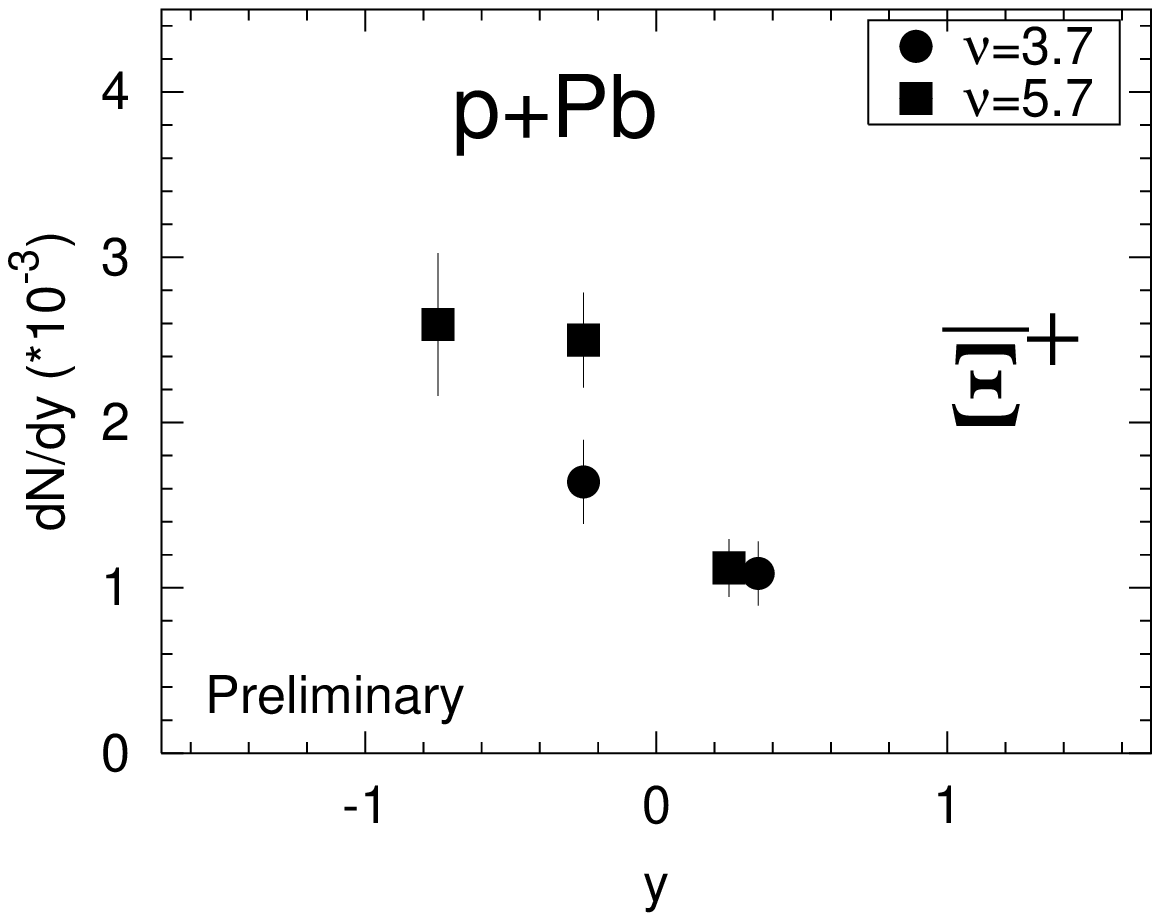,width=2.0in}
\caption{Cascade baryon yields in p+p ({\em left}) and in
centrality controlled p+Pb collisions ({\em middle and right})
\label{fig:xiyield}}
\end{figure}

	Results on cascade baryon production \cite{qmsusa} are shown
	in Figure \ref{fig:xiyield}. The p+Pb sample is divided into
	two subsets of different centrality, in which the mean number
	of collisions are found to be $3.7$ and $5.7$. The minimum
	bias p+Pb collision corresponds to $\nu=3.75$. The baryon
	stopping effect is visible resulting in the decrease of the
	forward yield in more central p+A collisions.

\subsection{Enhancement of Strange Baryons in Proton-Nucleus and Nucleus-Nucleus Interactions}

	The new measurements in p+p collision provide a new reference
	of elementary interaction for extrapolation to A+A
	collisions. Figure \ref{fig:xienh} compares the hyperon yields
	in p+p, p+A and Pb+Pb collisions at midrapidity, using the
	WA97 \cite{antinori} and NA49 results.

	The solid line on the figure corresponds to the simple Wounded
	Nucleon Model (WNM) prediction. An enhancement is visible in
	p+A, which means that the predicted scaling breaks already
	for p+A interactions. %The enhancement of baryons is higher
	%than that of the antibaryons, which can be due to stopping.

%\begin{figure}[!ht]
%\psfig{figure=enh_pA.ps,width=3.1in}
%\caption{Enhancement in pA (to be kept?) \label{fig:xienhpb}}
%\end{figure}

\subsection{Different Interpretations of the Enhancement}

\begin{figure}[!ht]
\begin{center}
\psfig{figure=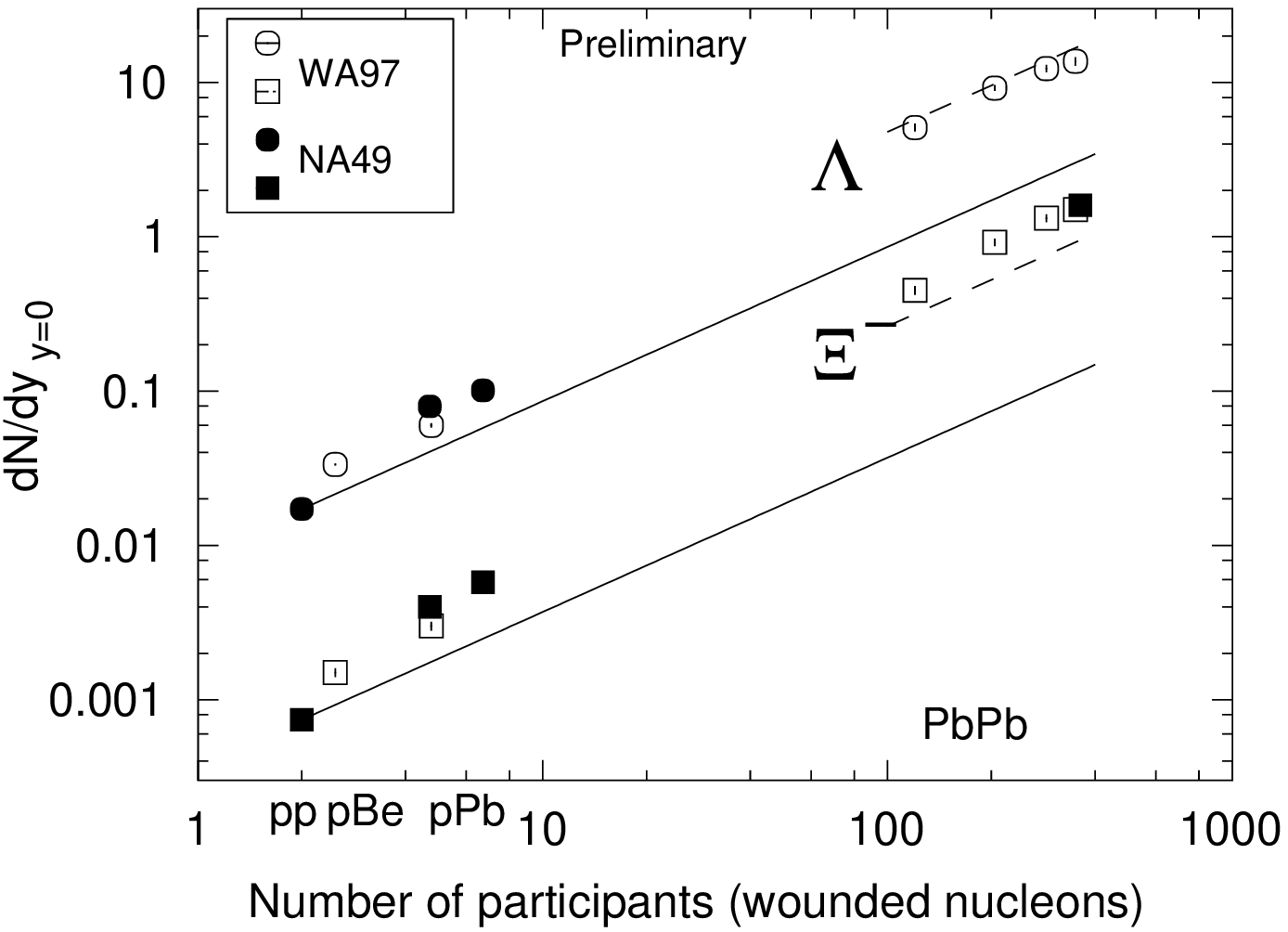,width=2.7in}
\psfig{figure=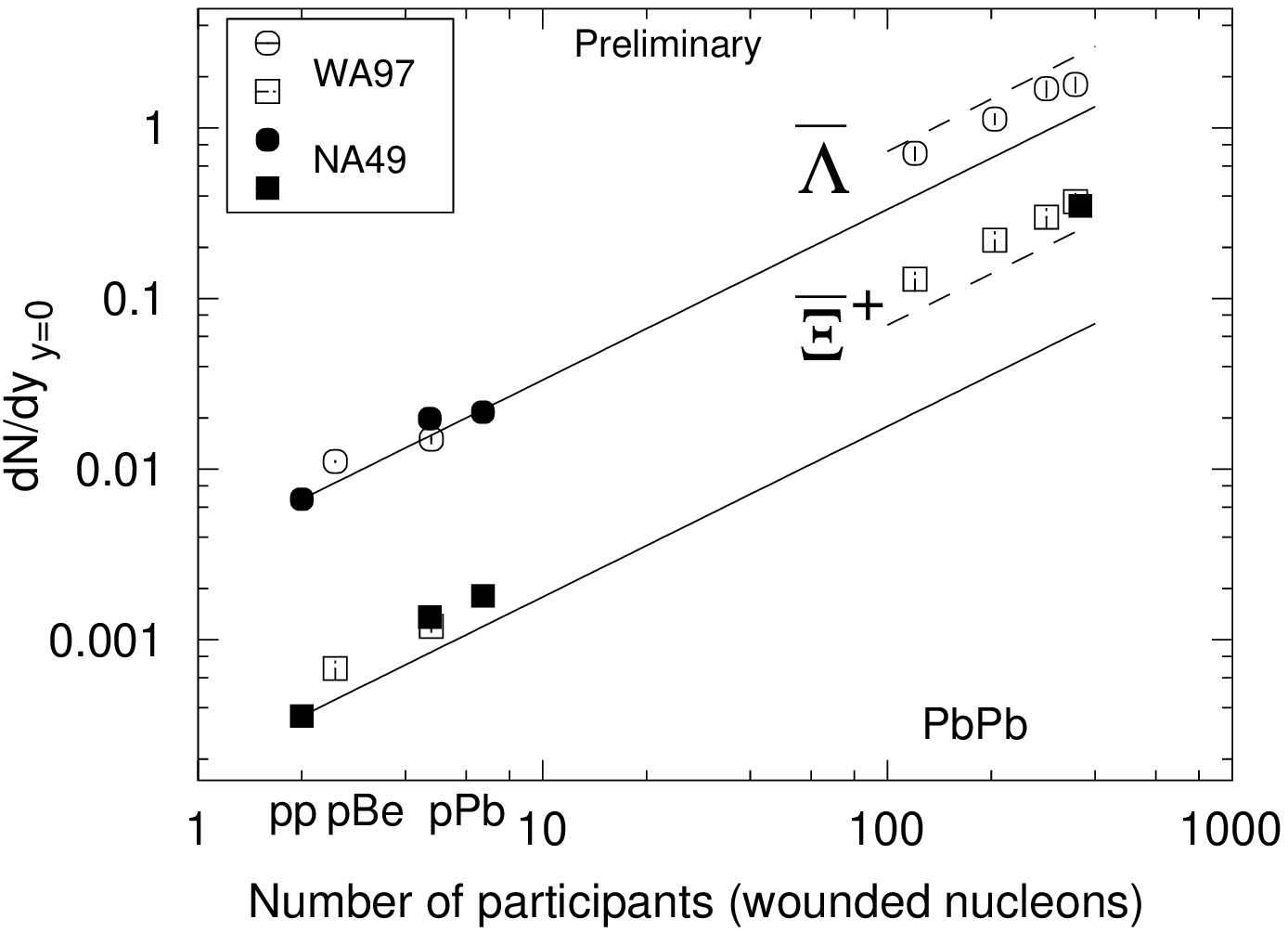,width=2.7in}
\caption{Midrapidity yields of hyperons in p+p, p+A and A+A collisions
as a function of $N_{part}$ \label{fig:xienh}}
\end{center}
\end{figure}

	The WNM is a two component picture: both the target and the
	projectile sides deliver particles to midrapidity. The simplest
	assumption is that the yield is proportional to $N_{part}$. In
	case of p+A, when the projectile has undergone $\nu$
	collisions, $N_{part}=\nu+1$. This predicted scaling is
	indicated on Figure \ref{fig:xienh} with the solid line, using
	p+p as a reference. This scaling apparently breaks for p+A
	interactions, so it can not be used for extrapolation to A+A.

	One can consider a simple modification \cite{qmsusa} of the
	WNM: let's assume that in case of p+A, the target side
	delivers $\nu$ times half of the p+p yield (as does the target
	side of p+p), all rest of the observed yield comes from
	projectile fragmentation. From the measurement, the
	contribution of the projectile side can be determined in p+A.
	This gives a prediction for A+A at the same $\nu$ assuming
	that both sides of A+A behave as the projectile side of
	p+A. This prediction is shown on Figure \ref{fig:xienh} with
	the dashed line.

	The difference between the conclusions of two oversimplified
	assumptions asks for a better understanding of p+A and A+A
	collisions. As the huge predicted increase of the yields
	compared to the WNM relied on a modest increase in p+A
	compared to p+p, precise measurements are also needed. Amongst
	others, one has to take into account the isospin-effect,
	arising from the neutron content of the nuclei; the question
	can be investigated by the analysis of last year's deuteron
	beam dataset. The other important effect is rescattering in
	the target, which can increase the target contribution.

%\subsection{Antibaryon-Baryon Ratio}

%\parbox[c]{2.6in}{\psfig{figure=ab_b_ratio.ps,width=2.6in}
%\footnotesize{\centerline{Figure ?: Antibaryon to baryon ratio}}}
%\parbox[c]{3.6in}{duma duma duma...}

%	Ratio goes down both in pp and pA; baryon increase due to stopping (?)

\section{Conclusions}

	The new measurement at 40 AGeV beam energy indicates a
	non-monotonic behaviour of the $K^+/\pi^+$ yield in A+A
	collisions as a function of the beam energy. The upcoming
	results at 20, 30 and 80 AGeV will provide additional
	information. The study of the system size dependence revealed
	that the variable $R-b/2$ allows a common scaling unlike $N_{part}$.

	The first results on cascade production in p+p collisions give
	a new reference for p+A and A+A collisions. The results show
	an enhancement in p+Pb compared to the simple WNM model, which
	makes the extrapolation to A+A collisions within this
	framework questionable.

\section*{Acknowledgements}

	I'd like to thank the Organisers for the pleasant and useful
	meeting. The work was partially supported by the Hungarian
	Scientific Research Foundation (F034707 and T32293).

\end{document}